\documentclass[aps,prb,twocolumn,showpacs,superscriptaddress]{revtex4-1}
\usepackage[latin1]{inputenc}
\usepackage{amsmath}
\usepackage{amssymb}
\usepackage{siunitx}
\usepackage[english]{babel}
\usepackage{graphicx, color}
\usepackage{hyperref}
\definecolor{link}{rgb}{0.1,0.1,0.9}
\hypersetup{colorlinks=true,linkcolor=link,citecolor=link,urlcolor=link,linktocpage}
\usepackage{epstopdf}
\usepackage{color}
\usepackage{longtable}

\begin{document}
\title{Rotating magnetocaloric effect in the ferromagnetic Weyl semi-metal Co$_{3}$Sn$_{2}$S$_{2}$}

\author{Anzar Ali$^*$, Shama$^*$, and Yogesh Singh}
\affiliation{Department of Physical Sciences, Indian Institute of Science Education and Research, Knowledge city, Sector 81, SAS Nagar, Manauli PO 140306, Mohali, Punjab, India}

\begin{abstract}
	
The rotating magnetocaloric effect (RMCE) is a recent interest in magnetic refrigeration technique in which the cooling effect is attained by rotating the anisotropic magnetocaloric material from one orientation to the other in a fixed magnetic field. In this work, we report the anisotropic magnetocaloric properties of single crystals of the ferromagnetic Weyl semimetal Co$_{3}$Sn$_{2}$S$_{2}$ for magnetic field $H\parallel c$ axis and $H\parallel ab$ plane.  We observed a significant (factor of $2$) difference between the magnetocaloric effect measured in both orientations. The rotating magnetocaloric effect has been extracted by taking the difference of the magnetic entropy change ($\Delta S_{M}$) for fields applied in the two crystallographic orientations.  In a scaling analysis of $\Delta S_{M}$, the rescaled $\Delta S_{M}(T,H)$  vs reduced temperature $\theta$ curves collapse onto a single universal curve, indicating that the transition from paramagnetic to ferromagnetic phase at 174~K is a second order transition. Furthermore, using the power law dependence of $\Delta S_{M}$ and relative cooling power RCP, the critical exponents $\beta$ and $\gamma$ are calculated, which are consistent with the recent critical behavior study on this compound \cite{Yan2018}.           

\end{abstract}

\maketitle

\section{Introduction}

The magnetocaloric effect (MCE) has been a topic of sustained attention for the last two decades due to its potential application in magnetic refrigeration \cite{Tegus, Spichkin, Franco, Ali}.  Providing an environmental friendly alternative to conventional gases is one of the main advantage of this technique. MCE quantifies the thermal response of a magnetic material caused by a varying magnetic field. The rotating magnetocaloric effect (RMCE) is a new direction in magnetic refrigeration that has drawn strong interest recently \cite{Balli, Zhang, Tkac, Lorusso, Oliveira, Jia, Liu, Konieczny}.  The conventional MCE utilises a change in the magnetic entropy on variation of a magnetic field.  In RMCE advantage can be taken of the anisotropy in a material's magnetocaloric response for magnetic fields in different crystallographic directions.  In RMCE, the change in magnetic entropy (-$\Delta S_{R}$) is obtained by rotating the magnetic sample from one crystallographic orientation to another, in a fixed external magnetic field.  RMCE is a potentially more favorable technique in comparison to MCE where cooling is achieved by changing the magnetic field. The other advantage of RMCE is the possible use of a permanent magnet or a fixed field which will lower the running cost.

There has been immense recent interest in the layered shandite-type half metallic ferromagnet Co$_{3}$Sn$_{2}$S$_{2}$ which shows a paramagnetic to ferromagnetic phase transition (PM-FM) at $T_c = 174$~K \cite{Holder, Kassem, Kassem2016}.  Co$_3$Sn$_2$S$_2$ is also a Weyl semi-metal and shows strong intrinsic anomalous Hall effect due to the presence of Weyl fermions near the Fermi energy \cite{Liu2018, Wang2016}.  Recently planar Hall effect has been reported on this material far below $T_c$, confirming the unconventional effects of the topological band structure \cite{Shama}. The crystal structure of Co$_{3}$Sn$_{2}$S$_{2}$ is built up of metallic hexagonal layers stacked along the crystallographic $c$-axis \cite{Sakai, Weihrich}. Such a layered structure may lead to large magnetocrystalline anisotropy \cite{Johnson}.  Thus, the half metallic ferromagnet Co$_{3}$Sn$_{2}$S$_{2}$ is a suitable material to study the RMCE.  There is also a push in the community to marry topological properties with other functionalities, and Co$_{3}$Sn$_{2}$S$_{2}$ would be a good candidate to look at MCE in a topological material.  Recently, a critical behavior study of this half metallic ferromagnet  Co$_{3}$Sn$_{2}$S$_{2}$ has been done by Yan \textit{et al.} \cite{Yan2018} and Shi \textit{et al.} have reported a magnetocaloric study on polycrystalline Co$_{3}$Sn$_{2}$S$_{2}$ \cite{Shi}.  However, no study of the anisotropic MCE and the potential of Co$_3$Sn$_2$S$_2$ as an RMCE material has been reported.

In this work, we have grown single crystals of Co$_3$Sn$_2$S$_2$ and measured the anisotropic magnetization as a function of temperature and as a function of magnetic field $H$.  The isothermal magnetization data are used to extract the anisotropic magnetocaloric effect ($-\Delta S_{M}$) with $H$ applied parallel to the $c$-axis ($H\parallel c$) and perpendicular to $c$-axis, that is in $ab$-plane ($H\parallel ab$).  It is found that the magnetization and the MCE is strongly anisotropic.  The RMCE (-$\Delta S_{R}$) is then calculated by taking the difference of ($-\Delta S_{M}$) obtained for the two crystallographic orientations.  We have further studied the nature of the paramagnetic to ferromagnetic phase transition by attempting to scale the MCE data at various $T$ and $H$.  The successful construction of a universal MCE curve indicates that the ferromagnetic transition is second order in nature.  Finally, we have calculated the power law exponents from fits of the magnetic field dependence of the maximum in $-\Delta S_{M}$ and the RCP (relative cooling potential) and related them to critical exponents for the transition.  We found that our values of the power law exponents are consistent with the recently reported critical exponents \cite{Yan2018}.           

\section{Experimental details}

Single crystals of Co$_{3}$Sn$_{2}$S$_{2}$ were grown as described recently \cite{Shama}.  All the magnetic measurement as a function of applied magnetic field and temperature have been performed using the vibrating sample magnetometer (VSM) option of a Quantum Design Physical Property Measurement System (PPMS). For the magnetocaloric study, anisotropic isothermal magnetization data has been collected in the temperature range $155$ to $195$~K around the ferromagnetic $T_c \approx 175$~K and in magnetic fields up to $9$~T for a Co$_{3}$Sn$_{2}$S$_{2}$ single crystal.

\begin{figure}
	\centering
	\includegraphics[width=1\linewidth, height=0.35\textheight]{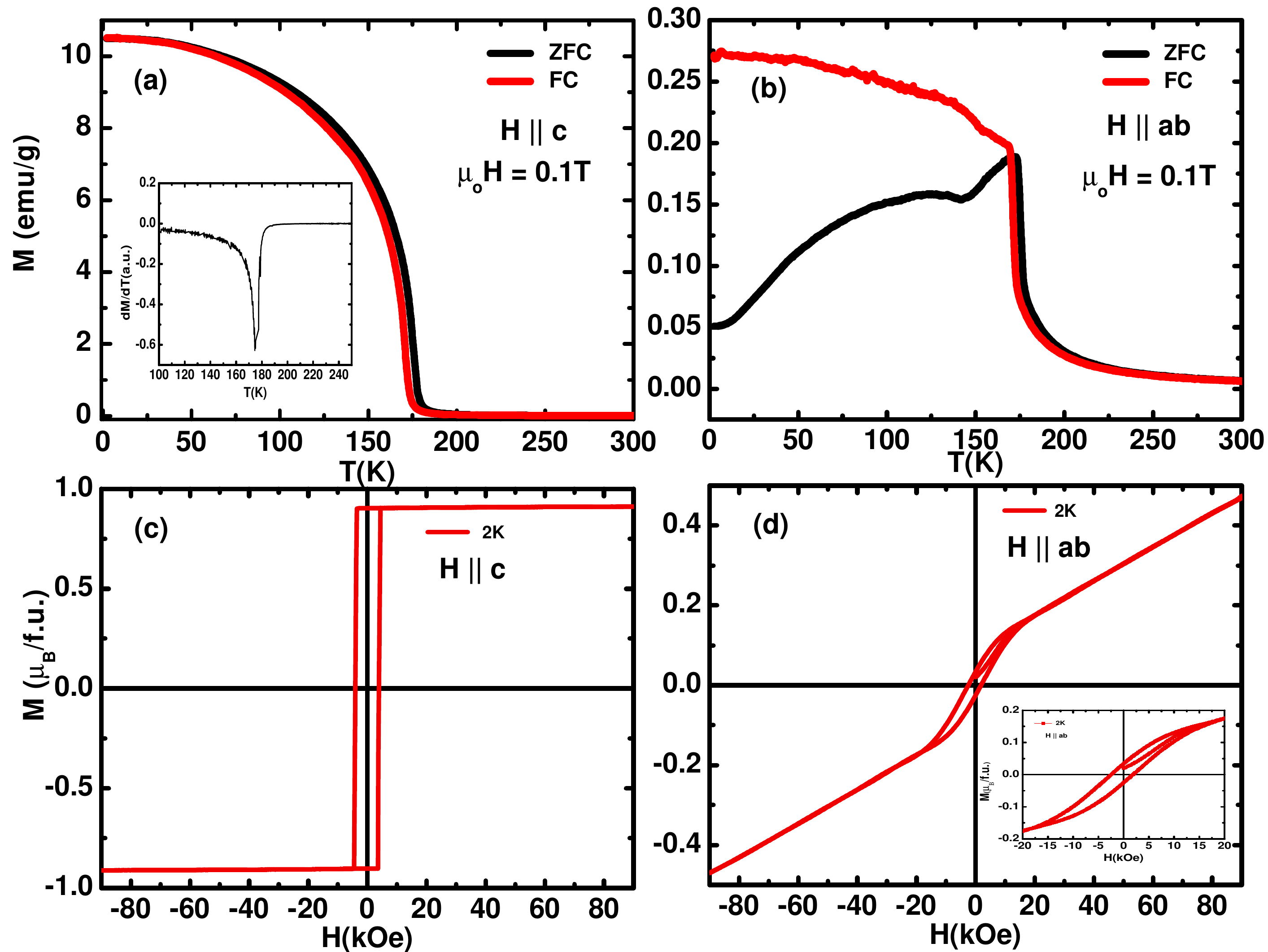}
	\caption{Temperature dependence of zero-field cooled (ZFC) and field cooled (FC) magnetization measured in a field of $0.1$~T applied along (a) the $c$-axis and (b) in the $ab$-plane.  (c) Magnetization $M$ at $2$~K along the $c$-axis, and (d) in the ab plane. Inset in (a) shows $dM/dT$ vs $T$ plot and inset of (d) shows the enlarged $M$-$H$ loop in low field highlighting the small hysteresis.}
	\label{Fig-M}
\end{figure}
  
\section{Results and Discussion}

Figures~\ref{Fig-M}(a) and (b) show the zero field cooled (ZFC) and field cooled (FC) magnetization $M$ vs temperature $T$ measured in a magnetic field $H = 0.1$~T applied along the $c$-axis and within the $ab$-plane, respectively.  The inset of Fig.~\ref{Fig-M}(a) shows the dM/dT vs T plot with a minima at $T_c \approx174$~K confirming the paramagnetic to ferromagnetic phase transition \cite{Schnelle}. The magnetization is strongly ansotropic, increasing to much larger values for $H\parallel c$ than for $H\parallel ab$.  The ZFC and FC curve for $H\parallel ab$ show a large bifurcation below $T_c$ which most likely originates from ferromagnetic domains. Figures~\ref{Fig-M}(c) and (d) show the field dependence of magnetization M(H) at 2~K for $H\parallel c$ and $H\parallel ab$, respectively. For the case of  $H\parallel c$ the magnetization increase abruptly and reaches saturation in a very low field with a narrow hysteresis loop, while for $H\parallel ab$ the magnetization, after an initial rapid increase, show a linear increase with field and does not saturate up to the highest fields measured $9$~T\@.  This indicates that the crystallographic $c$-axis is the magnetic easy axis.  Inset of Figure~\ref{Fig-M}(d) shows the hysteresis loop for small fields.  The small hysteresis in both directions indicate that Co$_{3}$Sn$_{2}$S$_{2}$ is a soft ferromagnet.

%The magnetization curves start to saturate below $T_c$ at a relatively small field for the $H\parallel c$ case, while it increase slowly for the case $H\parallel ab$ and hard to saturate in the applied field up to 9~T. This behavior is confirming the magnetic anisotropy and easy c-axis.

\begin{figure}
	\centering
	\includegraphics[width=1\linewidth, height=0.35\textheight]{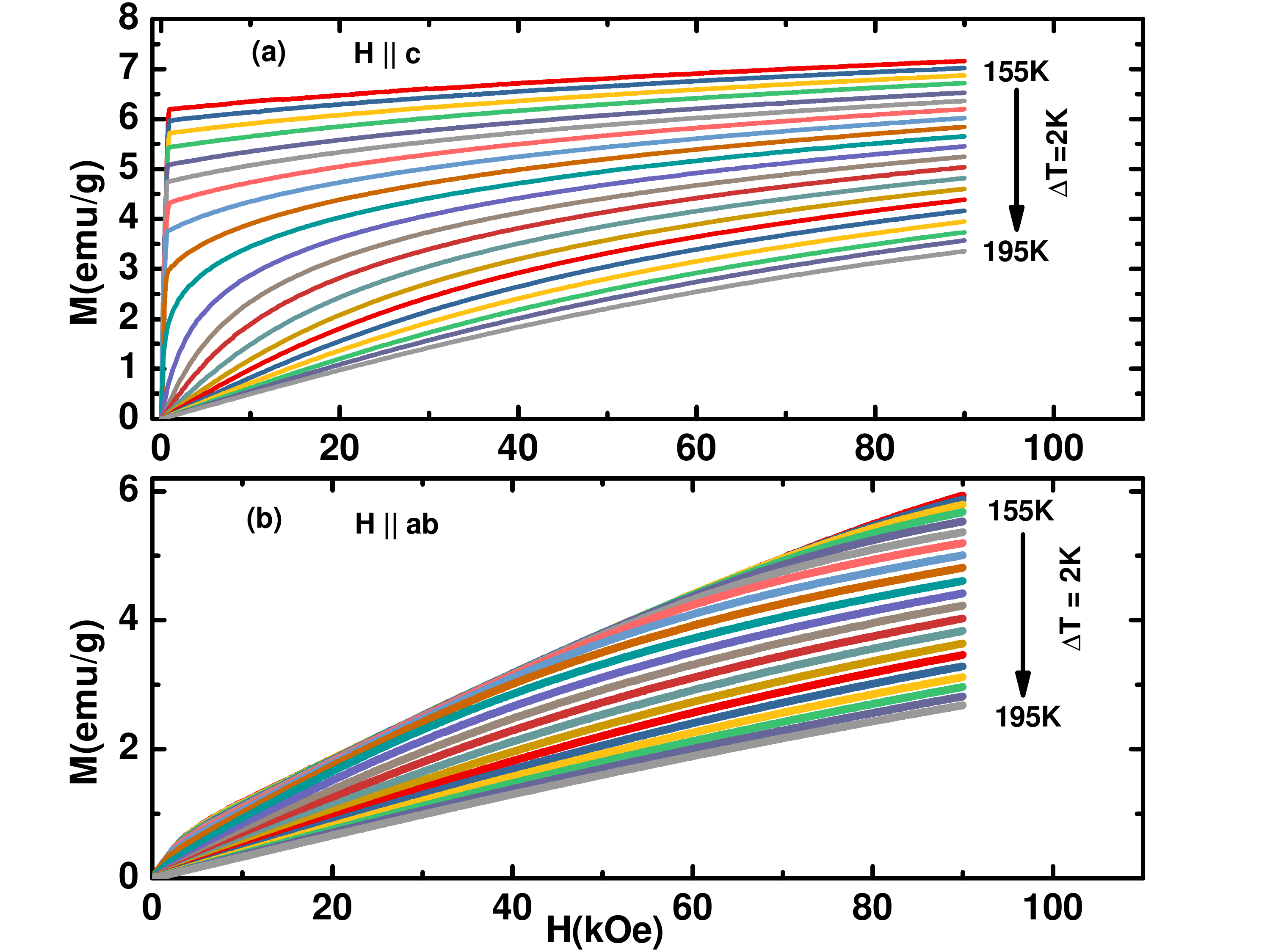}
	\caption{A series of magnetization isotherms at various temperatures with a magnetic field (a) along the $c$-axis (b) in the $ab$-plane.}
	\label{Isothermal}
\end{figure}

We now turn to the magnetocaloric response of Co$_3$Sn$_2$S$_2$, which we obtain from isothermal magnetization. Figures~\ref{Isothermal}(a) and (b) show a series of field dependent isothermal magnetization data recorded for temperatures around the critical temperature $T_c \approx 174$~K for the configurations $H\parallel c$ and $H\parallel ab$, respectively. The magnetization data was collected in the temperature range $155$~K to $195$~K at intervals of $2$~K\@.  The MCE (= negative magnetic entropy change $-\Delta S_M$) can be obtained using Maxwell's relation \cite{Ali},

\begin{equation}
\Delta S_M (T, H) = \mu_{0} \int _{0}^{H} \left[ \dfrac{\partial S(T,H)}{\partial T}\right]_T dH.
\label{Maxwell_1}
\end{equation}
Using thermodynamics relations, it can be further written as:
\begin{equation}
\Delta S_M (T, H) = \mu_{0} \int _{0}^{H} \left[  \dfrac{\partial M(T,H)}{\partial T}\right]_H dH.
\label{Maxwell_2}
\end{equation}
If magnetization data is available only at closely spaced discrete values of temperature, as is mostly the case in experiments, the integral in the above expression is replaced by a summation \cite{Ali}.

\begin{equation}
\Delta S_M (T, H) = \mu_{0} \sum_{i}^{} \dfrac { M(T_{i+1},H) -  M(T_i,H)}{T_{i+1} - T_i}.
\label{Maxwell_3}
\end{equation}
                 
\begin{figure}
\centering
\includegraphics[width=1\linewidth, height=0.6\textheight]{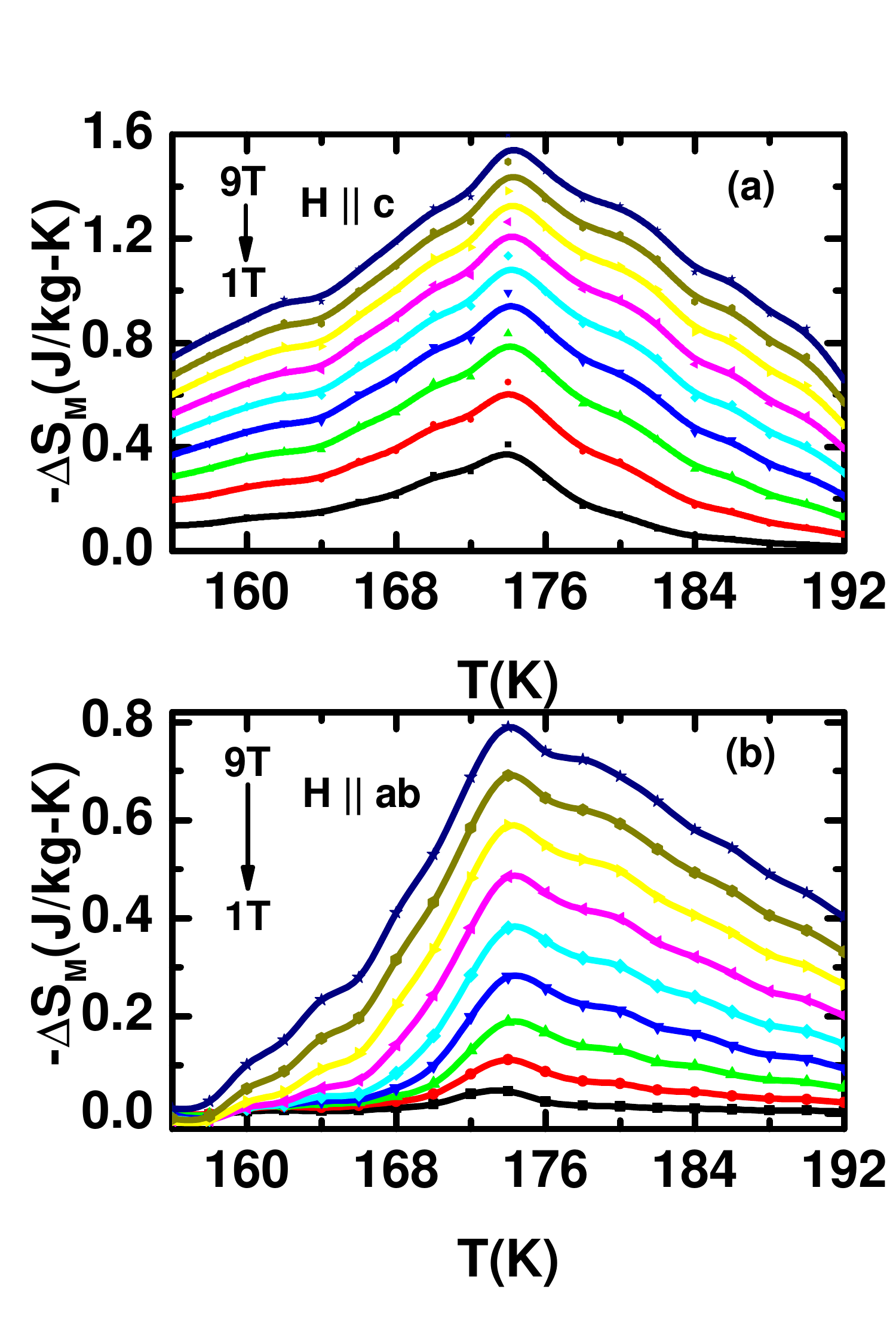}
\caption{Temperature dependence of magnetic entropy change (-$\Delta S_{M}$) at various magnetic fields applied (a) along the $c$-axis, and (b) in the $ab$-plane.}
\label{MCE}
\end{figure}

Figures~\ref{MCE}(a) and (b) display the calculated temperature dependence of -$\Delta S_{M}$ in  various fields up to 9T applied in both $H\parallel c$ and $H\parallel ab$ geometries respectively. The -$\Delta S_{M}$ vs T curves for both the cases show broad peaks around $T_c = 174$~K.  The maximum value of $-\Delta S_{M}$ at $9$~T is $1.6$~J/kg-K along the $c$-axis, while it is $0.8$~J/kg-K in the $ab$~plane. The large factor of $2$ anisotropy in the magnetic entropy change $-\Delta S_{M}$ between the two crystallographic orientations makes Co$_{3}$Sn$_{2}$S$_{2}$ a good candidate for the RMCE.  Additionally, $-\Delta S_{M}$ is negative at low temperature and low field for $H\parallel ab$, while it is positive along $c$-axis, indicating a sign change anisotropy which can also be potentially exploited.    

\begin{figure}
\centering
\includegraphics[width=1\linewidth, height=0.3\textheight]{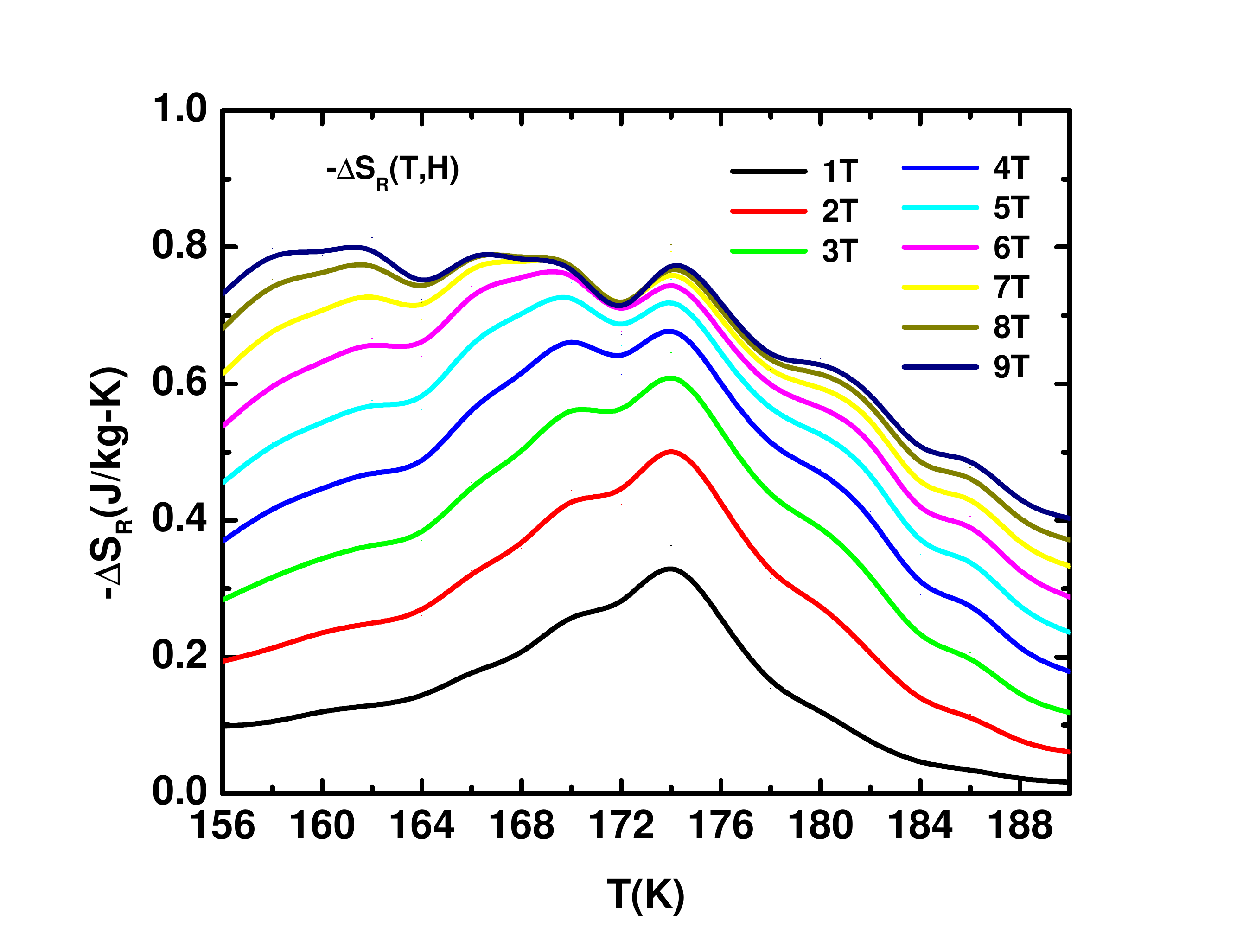}
\caption{Temperature dependence of the RMCE : magnetic entropy change -$ \Delta S_{R}$ calculated by rotating the crystal from $ab$-plane to the $c$-axis in various magnetic fields.}
\label{RMCE}
\end{figure}

The temperature and field dependent rotating magnetocaloric entropy change $\Delta S_R(T,H)$ can be calculated from the magnetic entropy change in the two crystallographic directions using the equation:

\begin{equation}
\begin{split}
\Delta S_R (T, H) = S_M (T,H||c ) - S_M (T, H||{ab} ) \\
                  = [S_M (T,H||c ) - S_M (T, 0 )] \\
                  - [S_M (T,H||{ab} ) - S_M (T, 0 )] \\
                  = \Delta S_M (T,H||c) - \Delta S_M (T, H||{ab})
\label{Rot}
\end{split}
\end{equation}

Figure~\ref{RMCE} shows the temperature dependence of the RMCE (-$ \Delta S_{R}$) calculated using equation(\ref{Rot}) in various applied magnetic fields as indicated in the plot. The maximum rotational entropy change (-$ \Delta S_{R}$) at $T_c$ increases with increasing field suggesting that the anisotropy of the magnetic entropy change between the two directions increases with field.  Additionally, the width of the peak also becomes broader on increasing magnetic field and it becomes approximately constant below $T_c$ for higher fields. Thus, for a broad temperature range near and below $T_c$, a significant magnetic entropy change can be achieved by rotating from the $ab$-plane to the $c$-axis in a fixed magnetic field.  This is an advantage over the conventional MCE where an entropy change can only be achieved by changing the magnetic field.
  
\begin{figure}
\centering
\includegraphics[width=1\linewidth, height=0.6\textheight]{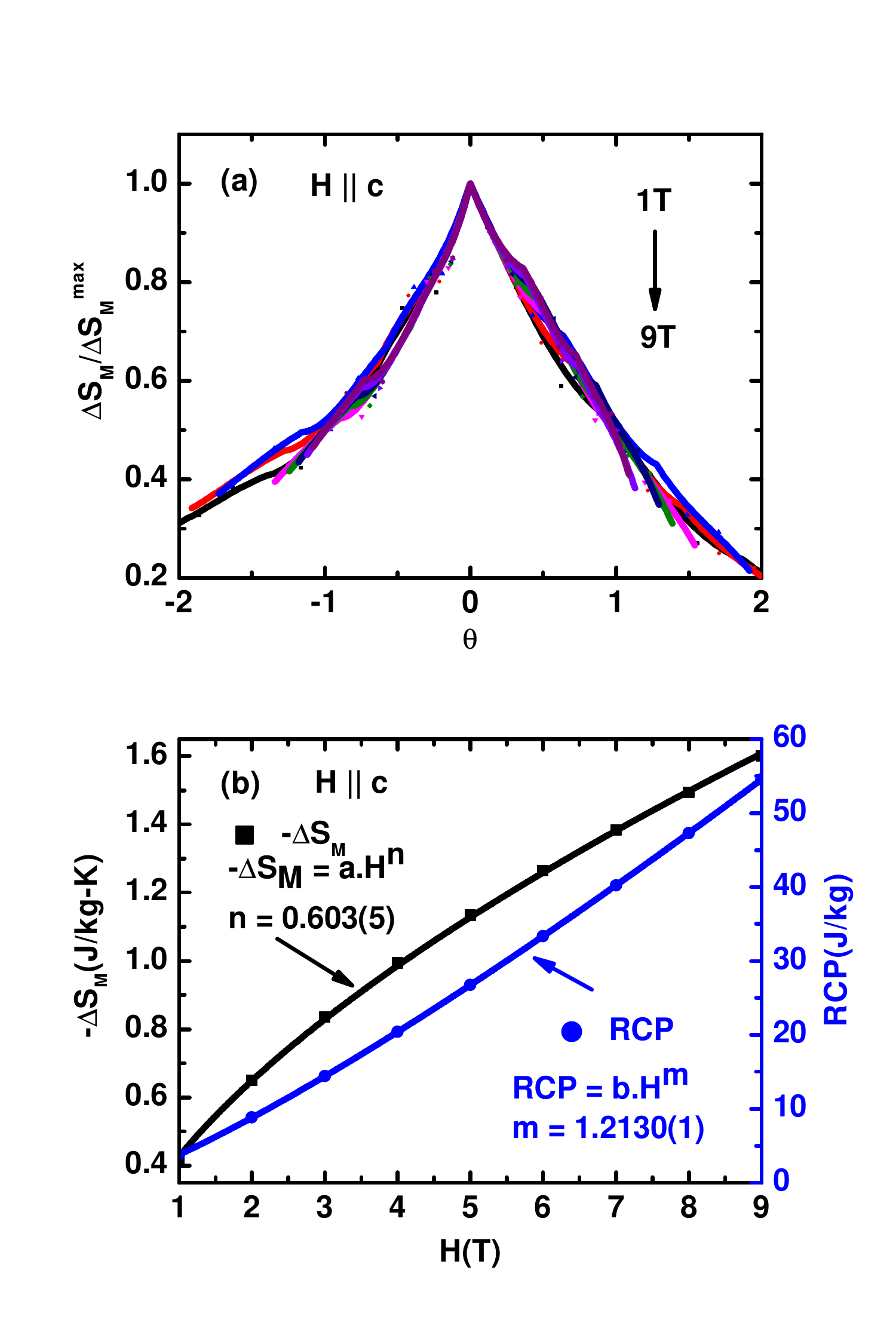}
\caption{(a) Universal magnetocaloric curve when plotted as rescaled entropy $\Delta S_{M}/\Delta S_M^{\rm max}$ vs reduced temperature $\theta$. (b) Power law fitting of -$\Delta S_{M}$ vs H and RCP vs H. }
\label{Scaling}
\end{figure}

In order to get further insight about the nature of the paramagnetic to ferromagnetic phase transition at $174$~K, we have attempted a scaling of the magnetocaloric data at various temperatures and fields using the entropy scaling laws reported recently by Franco \textit{et al.} \cite{Franco2010}. They proposed a phenomenological universal curve for the magnetic entropy change near a second order phase transition in which all the $\Delta S_M(T,H)$ data collapse onto a single universal curve when plotted as the normalized entropy ($\Delta S_M/\Delta S_M^{\rm max}$) at each field vs a reduced temperature $\theta$ given by:
\begin{equation}
\theta = -{T - T_c \over T_{r1} - T_c} \;\;\; \textrm{for}\;\; \; T \leq T_c
\end{equation}
\begin{equation}
\theta = {T - T_c \over T_{r2} - T_c}  \;\;\;\; \textrm{for}\;\; \; T > T_c~,
\end{equation}
\\
where $T_{r1}$ and $T_{r2}$ are the temperatures at the full width at half maximum of the anomaly in $\Delta S_M$ at $T_c$.  Figure~\ref{Scaling}(a) show that all the $\Delta S_M$ curves for the different magnetic fields approximately collapse onto a single universal master curve.  This is strong evidence of the second order nature of the ferromagnetic transition at $T_c = 174$~K in Co$_{3}$Sn$_{2}$S$_{2}$.

For materials having a second order magnetic phase transition, the field dependence of the maximum in $-\Delta S_{M}$ is expected to follow a power law field dependence \cite{Franco2006}:
\begin{equation}
-\Delta S_{M}^{\rm max} = aH^{n}
\end{equation} 

Figure~\ref{Scaling}(b) shows results of such a power law fitting to the $-\Delta S_{M}^{\rm max}$ vs $H$ data.  The fit gave the value for the power law exponent $n = 0.603(5)$.  The expected value for $n$ within a Mean field model is $2/3$ \cite{Oesterreicher}. The deviation of the value of $n$ from $2/3$ indicates that the transition from paramagnetic to ferromagnetic phase at $T_c$ deviates from expectations for a mean field transition. The parameter $n$ can be relate to the critical exponents of a second-order transition by the expression:
\begin{equation}
 n = 1 + \dfrac{\beta - 1}{\beta + \gamma},
\end{equation}
\noindent
where $\beta$ is the critical exponent extracted from the spontaneous magnetization below $T_c$, and $\gamma$ is the critical exponent extracted from the inverse susceptibility above $T_c$.

The recent critical behavior study of the ferromagnetic transition in polycrystals of Co$_3$Sn$_2$S$_2$ by Yan \textit{et al.} \cite{Yan2018} gave the following values of exponents: $\beta = 0.356$ and $\gamma = 1.26$.  Inserting these values in the above equation, we obtain $n = 0.601$ which is in good agreement with the value of $n = 0.603$ we obtained above by a power law fitting of the magnetic entropy data.  This further supports the second-order nature of the phase transition. 

The relative cooling power (RCP), which quantifies the usefulness of a magnetocaloric (MC) material, is defined as the product of the full width at half maximum $\Delta T_{\rm FWHM}$ and $ \Delta S_M^{\rm max}$. RCP also depends on the magnetic field and is expected to have a power law field dependence RCP = bH$^{m}$, where $m$ can be related to the critical exponent  $\delta$, obtained from the isothermal magnetization at $T_c$, by the relation \cite{Franco2008}:

\begin{equation}
m = 1 + \dfrac{1}{\delta}
\end{equation}

Figure~\ref{Scaling}(b) show the RCP vs magnetic field.  From the power law fitting, we obtain the value $ m = 1.213$, which gives the value of $\delta = 4.695$, which is close to the value estimated from a previous Arrott plots analysis \cite{Yan2018}. So the construction of universal curve and power law fitting for maximum entropy change and RCP gives further insight about critical behavior of the ferromagnetic phase transition in Co$_{3}$Sn$_{2}$S$_{2}$. 
  
\section{Conclusion}
Co$_{3}$Sn$_{2}$S$_{2}$ is a half-metallic ferromagnet and a Weyl semi-metal candidate showing a paramagnetic to ferromagnetic phase transition at $T_c = 174$~K.  Measurements on single crystals of Co$_3$Sn$_2$S$_2$ reveal a large anisotropy in magnetic properties which can be exploited for potential applications.
In this work, we report the anisotropic magnetocaloric (MCE) response -$\Delta S_{M} $ of single crystals of Co$_{3}$Sn$_{2}$S$_{2}$ for $H\parallel c$ axis and $H\parallel ab$ plane near the PM-FM transition. The maximum entropy change (-$\Delta S_{M}$) is about $1.6$~J/kg-K for $H\parallel c$ while the maximum entropy change for $H\parallel ab$ plane is $0.8$~J/kg-K\@.  This factor of $2$ anisotropy in $-\Delta S_M$ in the two crystallographic directions make Co$_3$Sn$_2$S$_2$ a good candidate to exploit the rotating MCE in which a large magnetic entropy change can be achieved at a fixed magnetic field by just rotating the crystal.  To gain further insight into the nature of the ferromagnetic transition we have performed a scaling analysis of the magnetic entropy change at various temperature and fields.  The successful construction of a universal magnetocaloric curve indicate that the PM-FM transition is a second order magnetic transition. Powr law dependencies of the magnetic entropy change and RCP were found and yield the power law exponents $n = 0.603(5)$ and $m = 1.213(12)$, respectively.  These values are consistent with values of the critical exponents reported in a previous critical behavior study on polycrystalline Co$_{3}$Sn$_{2}$S$_{2}$. 
 
\section{Acknowledgment}
We acknowledge use of the x-ray facility and the SEM facility at IISER Mohali.  \\\\
\noindent
$^*$ These authors contributed equally to this work.
%\bibliographystyle{apsrev4-1}
%\bibliography{Ref}

\end{document}